\documentclass[aps,prl,reprint]{revtex4-2}
\usepackage{amsmath}
\usepackage{amsfonts}
\usepackage{xfrac} %
\usepackage{bbm}
\usepackage{tikzit}
\usepackage{xfrac}

\tikzstyle{inline text}=[text height=1.5ex, text depth=0.25ex, yshift=0.5mm]
\tikzstyle{upground}=[circuit ee IEC, thick, ground, rotate=90, scale=2]
\tikzstyle{downground}=[circuit ee IEC, thick, ground, rotate=-90, scale=1.5]
\tikzstyle{point}=[regular polygon, regular polygon sides=3, draw, scale=0.75, inner sep=-0.5pt, minimum width=9mm, fill=white, regular polygon rotate=180, tikzit fill={rgb,255: red,242; green,255; blue,92}]
\tikzstyle{wide copoint}=[fill=white, draw, shape=isosceles triangle, shape border rotate=90, isosceles triangle stretches=true, inner sep=0pt, minimum width=1.5cm, minimum height=6.12mm]
\tikzstyle{wide point}=[fill=white, draw, shape=isosceles triangle, shape border rotate=-90, isosceles triangle stretches=true, inner sep=0pt, minimum width=1.5cm, minimum height=6.12mm, yshift=-0.0mm]
\tikzstyle{wide dpoint}=[wide point, doubled]
\tikzstyle{copoint}=[regular polygon, regular polygon sides=3, draw, scale=0.75, inner sep=-0.5pt, minimum width=9mm, fill=white, tikzit fill={rgb,255: red,255; green,128; blue,0}, tikzit draw={rgb,255: red,255; green,128; blue,0}]
\tikzstyle{dot}=[inner sep=0mm, minimum width=1mm, minimum height=1mm, draw, shape=circle]
\tikzstyle{black dot}=[dot, fill={gray!30}, text depth=-0.2mm]
\tikzstyle{white dot}=[dot, fill=white, text depth=-0.2mm]
\tikzstyle{small box}=[rectangle, inline text, fill=white, draw, minimum height=5mm, yshift=-0.5mm, minimum width=5mm, font={\small}]
\tikzstyle{small gray box}=[small box, fill={gray!30}]
\tikzstyle{medium box}=[rectangle, inline text, fill=white, draw, minimum height=5mm, yshift=-0.5mm, minimum width=10mm, font={\small}]
\tikzstyle{square box}=[small box]
\tikzstyle{medium gray box}=[small box, fill={gray!30}]
\tikzstyle{semilarge box}=[rectangle, inline text, fill=white, draw, minimum height=5mm, yshift=-0.5mm, minimum width=12.5mm, font={\small}]
\tikzstyle{large box}=[rectangle, inline text, fill=white, draw, minimum height=5mm, yshift=-0.5mm, minimum width=15mm, font={\small}]
\tikzstyle{large gray box}=[small box, fill={gray!30}]
\tikzstyle{dpoint}=[point, doubled]
\tikzstyle{dcopoint}=[copoint, doubled]
\tikzstyle{boldedge}=[doubled, shorten <=-0.17mm, shorten >=-0.17mm]
\tikzstyle{normal}=[line width=0.9pt]
\tikzstyle{doubled}=[line width=1pt]
\tikzstyle{boldedge}=[doubled, shorten <=-0.17mm, shorten >=-0.17mm]
\tikzstyle{small dbox}=[small box, doubled]
\tikzstyle{white ddot}=[white dot, doubled]
\tikzstyle{black ddot}=[black dot, doubled, tikzit fill=black]
\tikzstyle{map}=[draw, shape=NEbox, inner sep=2pt, minimum height=6mm, fill=white]
\tikzstyle{box}=[draw, shape=rectangle, inner sep=2pt, minimum height=6mm, minimum width=6mm, fill=white]
\tikzstyle{dbox}=[draw, doubled, shape=rectangle, inner sep=2pt, minimum height=6mm, minimum width=6mm, fill=white]
\tikzstyle{dmap}=[draw, doubled, shape=NEbox, inner sep=2pt, minimum height=6mm, fill=white]
\tikzstyle{dmapdag}=[draw, doubled, shape=SEbox, inner sep=2pt, minimum height=6mm, fill=white]
\tikzstyle{dmapadj}=[draw, doubled, shape=SEbox, inner sep=2pt, minimum height=6mm, fill=white]
\tikzstyle{dmaptrans}=[draw, doubled, shape=SWbox, inner sep=2pt, minimum height=6mm, fill=white]
\tikzstyle{dmapconj}=[draw, doubled, shape=NWbox, inner sep=2pt, minimum height=6mm, fill=white]
\tikzstyle{map}=[draw, shape=NEbox, inner sep=2pt, minimum height=6mm, fill=white]
\tikzstyle{dashedmap}=[draw, dashed, shape=NEbox, inner sep=2pt, minimum height=6mm, fill=white]
\tikzstyle{mapdag}=[draw, shape=SEbox, inner sep=2pt, minimum height=6mm, fill=white]
\tikzstyle{mapadj}=[draw, shape=SEbox, inner sep=2pt, minimum height=6mm, fill=white]
\tikzstyle{maptrans}=[draw, shape=SWbox, inner sep=2pt, minimum height=6mm, fill=white]
\tikzstyle{mapconj}=[draw, shape=NWbox, inner sep=2pt, minimum height=6mm, fill=white]
\tikzstyle{semilarge map}=[draw, shape=NEbox, inner sep=2pt, minimum height=6mm, fill=white, minimum width=9.5mm]
\tikzstyle{semilarge dmap}=[draw, doubled, shape=NEbox, inner sep=2pt, minimum height=6mm, fill=white, minimum width=9.5mm]
\tikzstyle{kpointdag}=[kpoint adjoint]
\tikzstyle{kpointadj}=[kpoint adjoint]
\tikzstyle{kpointconj}=[kpoint conjugate]
\tikzstyle{kpointtrans}=[kpoint transpose]
\tikzstyle{kpoint common}=[draw, fill=white, inner sep=1pt, minimum height=4mm]
\tikzstyle{kpoint sc}=[shape=cornerpoint, kpoint common]
\tikzstyle{kpoint adjoint sc}=[shape=cornercopoint, kpoint common]
\tikzstyle{kpoint}=[shape=cornerpoint, shorten left=5pt, kpoint common, tikzit fill={rgb,255: red,255; green,128; blue,0}]
\tikzstyle{kpoint adjoint}=[shape=cornercopoint, shorten left=5pt, kpoint common, tikzit fill={rgb,255: red,255; green,128; blue,0}]
\tikzstyle{kpoint conjugate}=[shape=cornerpoint, shorten right=5pt, kpoint common]
\tikzstyle{kpoint transpose}=[shape=cornercopoint, shorten right=5pt, kpoint common]
\tikzstyle{kpoint symm}=[shape=cornerpoint, shorten left=5pt, shorten right=5pt, kpoint common]
\tikzstyle{wide kpoint}=[kpoint, minimum width=1 cm, inner sep=2pt]
\tikzstyle{wide kpointdag}=[kpointdag, minimum width=1 cm, inner sep=2pt]
\tikzstyle{wide kpointconj}=[kpointconj, minimum width=1 cm, inner sep=2pt]
\tikzstyle{wide kpointtrans}=[kpointtrans, minimum width=1 cm, inner sep=2pt]
\tikzstyle{wider kpoint}=[kpoint, minimum width=1.25 cm, inner sep=2pt]
\tikzstyle{wider kpointdag}=[kpointdag, minimum width=1.25 cm, inner sep=2pt]
\tikzstyle{wider kpointconj}=[kpointconj, minimum width=1.25 cm, inner sep=2pt]
\tikzstyle{wider kpointtrans}=[kpointtrans, minimum width=1.25 cm, inner sep=2pt]
\tikzstyle{dkpoint}=[kpoint, doubled, tikzit fill={rgb,255: red,255; green,85; blue,210}]
\tikzstyle{wide dkpoint}=[wide kpoint, doubled, tikzit fill={rgb,255: red,68; green,255; blue,0}]
\tikzstyle{dkpointdag}=[kpoint adjoint, doubled]
\tikzstyle{wide dkpointdag}=[wide kpointdag, doubled]
\tikzstyle{label}=[fill=white, draw=white, shape=circle, tikzit draw={rgb,255: red,10; green,26; blue,255}, tikzit fill={rgb,255: red,0; green,12; blue,255}, font={\small}]
\tikzstyle{squarelabel}=[fill=white, draw=white, shape=rectangle, tikzit draw=black]
\tikzstyle{eslabel}=[tikzit draw={rgb,255: red,255; green,191; blue,191}, tikzit fill={rgb,255: red,255; green,191; blue,191}, font={\tiny}]
\tikzstyle{large dmap}=[draw, doubled, shape=NEbox, inner sep=2pt, minimum height=6mm, fill=white, minimum width=12mm]
\tikzstyle{gray point}=[point, fill={gray!40!white}]
\tikzstyle{gray dpoint}=[gray point, doubled, tikzit draw={rgb,255: red,128; green,128; blue,128}, tikzit fill={rgb,255: red,128; green,128; blue,128}]
\tikzstyle{gray copoint}=[copoint, fill={gray!40!white}, tikzit fill={rgb,255: red,128; green,128; blue,128}]
\tikzstyle{gray dcopoint}=[gray copoint, doubled, tikzit fill={rgb,255: red,128; green,128; blue,128}]
\tikzstyle{circlenew}=[draw=black, shape=circle, inner sep=1pt]
\tikzstyle{blue label}=[text=NavyBlue, tikzit draw={rgb,255: red,0; green,96; blue,167}, tikzit fill={rgb,255: red,35; green,68; blue,255}]
\tikzstyle{big dot}=[fill=white, draw=black, shape=circle, minimum width=6mm, minimum height=6mm]
\tikzstyle{3d box}=[fill=white, draw=black, shape=trapezium, trapezium left angle=-70, trapezium right angle=70, rotate=10]
\tikzstyle{slant red box}=[fill={rgb,255: red,191; green,0; blue,64}, draw={rgb,255: red,191; green,0; blue,64}, shape=rectangle, xslant=0.5, font={\tiny}, text={rgb,255: red,191; green,0; blue,64}, fill opacity=0.5, line width=1pt]
\tikzstyle{slant point}=[regular polygon, regular polygon sides=3, draw, scale=0.75, inner sep=-0.5pt, minimum width=9mm, fill white, regular polygon rotate=180, yslant=-0.3]
\tikzstyle{tiny orange label}=[font={\tiny}, text={rgb,255: red,255; green,128; blue,0}, tikzit draw={rgb,255: red,255; green,128; blue,0}]
\tikzstyle{tiny red label}=[font={\tiny}, text={rgb,255: red,191; green,0; blue,64}, tikzit draw={rgb,255: red,191; green,0; blue,64}, draw=none]
\tikzstyle{red label}=[text={rgb,255: red,191; green,0; blue,64}, tikzit draw={rgb,255: red,191; green,0; blue,64}]
\tikzstyle{slant label black}=[font={\tiny}, xslant=0.5, tikzit draw=black]
\tikzstyle{slant label red}=[font={\tiny}, xslant=0.5, text={rgb,255: red,191; green,0; blue,64}, tikzit draw={rgb,255: red,191; green,0; blue,64}]
\tikzstyle{slant label orange}=[font={\tiny}, xslant=0.5, text={rgb,255: red,255; green,128; blue,0}, tikzit draw={rgb,255: red,255; green,128; blue,0}]
\tikzstyle{slanted point}=[fill={rgb,255: red,191; green,0; blue,64}, draw={rgb,255: red,191; green,0; blue,64}, shape=triangle, regular polygon, regular polygon sides=3, scale=0.75, inner sep=-0.5pt, minimum width=5mm, regular polygon rotate=90, xslant=0.5, fill opacity=0.5, font={\tiny}, line width=1pt, text={rgb,255: red,191; green,0; blue,64}]
\tikzstyle{slanted point black}=[draw=black, shape=triangle, regular polygon, regular polygon sides=3, scale=0.75, inner sep=-0.5pt, minimum width=5mm, regular polygon rotate=90, xslant=0.5, font={\tiny}, line width=0.2pt, text=black, fill=white, tikzit fill=white]
\tikzstyle{red dot}=[fill={rgb,255: red,191; green,0; blue,64}, draw={rgb,255: red,191; green,0; blue,64}, shape=circle, inner sep=0, minimum width=1mm, minimum height=1mm]
\tikzstyle{black dot}=[fill=black, draw=black, shape=circle, inner sep=0, minimum width=1.5mm, minimum height=1.5mm]
\tikzstyle{orange dot}=[fill={rgb,255: red,255; green,128; blue,0}, draw={rgb,255: red,255; green,128; blue,0}, shape=circle, inner sep=0, minimum width=1.5mm, minimum height=1.5mm]
\tikzstyle{blue dot}=[fill={rgb,255: red,0; green,0; blue,228}, draw={rgb,255: red,0; green,0; blue,228}, shape=circle, inner sep=0, minimum width=1.5mm, minimum height=1.5mm]
\tikzstyle{slant white}=[fill=white, draw=black, shape=rectangle, xslant=0.5, font={\tiny}, line width=1pt]
\tikzstyle{slant small map}=[fill=white, draw=black, xslant=0.5, shape=rectangle, font={\tiny}, line width=1pt, inner sep=0.6mm]
\tikzstyle{slanted copoint black}=[draw=black, shape=triangle, regular polygon, regular polygon sides=3, scale=0.75, inner sep=-0.5pt, minimum width=5mm, regular polygon rotate=-90, xslant=0.5, font={\tiny}, line width=0.2pt, text=black, fill=white, tikzit fill=white]
\tikzstyle{purple dot}=[fill={rgb,255: red,128; green,0; blue,128}, draw={rgb,255: red,128; green,0; blue,128}, shape=circle, inner sep=0, minimum width=1.5mm, minimum height=1.5mm]
\tikzstyle{white dot 2}=[fill=white, draw=black, shape=circle]
\tikzstyle{horizontal point}=[style=point, rotate=-90, tikzit shape=rectangle, tikzit fill={rgb,255: red,191; green,128; blue,64}]
\tikzstyle{pslant orange}=[style=slanted point black, fill={rgb,255: red,255; green,128; blue,0}, draw={rgb,255: red,255; green,128; blue,0}, tikzit fill={rgb,255: red,255; green,128; blue,0}, tikzit draw={rgb,255: red,255; green,128; blue,0}]
\tikzstyle{upground horizontal}=[style=upground, rotate=-90]
\tikzstyle{double horizontal point}=[style=horizontal point, line width=1pt]
\tikzstyle{double point}=[style=point, line width=1pt]
\tikzstyle{double copoint}=[style=copoint, line width=1pt]
\tikzstyle{horizontal copoint}=[style=double copoint, rotate=-90]
\tikzstyle{slant label purple}=[style=slant label black, tikzit draw={rgb,255: red,128; green,0; blue,128}, text={rgb,255: red,128; green,0; blue,128}]
\tikzstyle{orange copoint}=[style=pslant orange, rotate=-180, tikzit fill={rgb,255: red,255; green,128; blue,0}]
\tikzstyle{new style 0}=[style=slant white, draw={rgb,255: red,0; green,0; blue,228}, fill={rgb,255: red,0; green,0; blue,228}, fill opacity=0.5, shape=rectangle]
\tikzstyle{wide slanted point}=[style=wide point, xslant=0.5, fill=white, rotate=-90, minimum width=0.8cm, fill={rgb,255: red,128; green,128; blue,128}, fill opacity=0.5, line width=1pt]
\tikzstyle{black dot white}=[style=black dot, text=white, draw=none, tikzit draw={rgb,255: red,191; green,255; blue,0}, shape=circle]

\tikzstyle{new edge style 1}=[-, line width=1pt, shorten <=-0.17mm, shorten >=-0.17mm, tikzit draw={rgb,255: red,204; green,0; blue,3}]
\tikzstyle{diredge}=[-, postaction=decorate, decoration={markings, mark=at position 0.55 with \edgearrow}]
\tikzstyle{bold diredge}=[-, diredge, line width=1pt, tikzit draw={rgb,255: red,128; green,0; blue,128}]
\tikzstyle{grey}=[-, draw={rgb,255: red,188; green,188; blue,188}]
\tikzstyle{classical}=[-, dashed, tikzit draw={rgb,255: red,255; green,128; blue,0}]
\tikzstyle{reddashed}=[-, dashed, draw={rgb,255: red,0; green,128; blue,128}, postaction=decorate, decoration={markings, mark=at position 0.55 with \edgearrow}]
\tikzstyle{reddahednoarrow}=[-, dashed, draw={rgb,255: red,179; green,40; blue,40}]
\tikzstyle{arrow edge}=[-, ->, draw={rgb,255: red,191; green,191; blue,191}, tikzit draw={rgb,255: red,191; green,191; blue,191}, ultra thick]
\tikzstyle{tarrow edge}=[-, ->, draw={rgb,255: red,191; green,191; blue,191}, tikzit draw={rgb,255: red,191; green,191; blue,191}]
\tikzstyle{gray edge}=[-, draw={rgb,255: red,191; green,191; blue,191}, tikzit draw={rgb,255: red,191; green,191; blue,191}, ultra thick]
\tikzstyle{lightgrayedge}=[-, draw={rgb,255: red,207; green,207; blue,207}]
\tikzstyle{green edge}=[-, tikzit draw={rgb,255: red,128; green,128; blue,0}, draw={rgb,255: red,128; green,128; blue,0}]
\tikzstyle{red edge}=[-, draw={rgb,255: red,191; green,0; blue,64}, tikzit draw={rgb,255: red,191; green,0; blue,64}]
\tikzstyle{arrow edge black}=[-, ->]
\tikzstyle{solid blue}=[-, draw={rgb,255: red,0; green,96; blue,167}, tikzit draw={rgb,255: red,0; green,96; blue,167}]
\tikzstyle{classical blue}=[-, draw={rgb,255: red,0; green,96; blue,167}, tikzit draw={rgb,255: red,0; green,96; blue,167}, dashed]
\tikzstyle{fill gray}=[-, fill=gray]
\tikzstyle{bold gray}=[-, line width=1pt, tikzit draw={rgb,255: red,128; green,128; blue,128}]
\tikzstyle{fill pink}=[-, fill={rgb,255: red,193; green,100; blue,94}, fill opacity=0.5, draw={rgb,255: red,134; green,68; blue,65}, line width=1pt, tikzit draw={rgb,255: red,134; green,68; blue,65}, tikzit fill={rgb,255: red,193; green,100; blue,94}]
\tikzstyle{fill carta da zucchero}=[-, fill={rgb,255: red,129; green,158; blue,219}, fill opacity=0.5, line width=0.4mm]
\tikzstyle{fill white}=[-, fill=white]
\tikzstyle{fill purple}=[-, fill={rgb,255: red,113; green,69; blue,128}, fill opacity=0.5, draw={rgb,255: red,79; green,48; blue,90}, tikzit fill={rgb,255: red,113; green,69; blue,128}, tikzit draw={rgb,255: red,79; green,48; blue,90}, line width=1pt]
\tikzstyle{fill green}=[-, fill={rgb,255: red,62; green,128; blue,120}, fill opacity=0.5, draw={rgb,255: red,33; green,68; blue,63}, tikzit fill={rgb,255: red,62; green,128; blue,120}, tikzit draw={rgb,255: red,33; green,68; blue,63}, line width=1pt]
\tikzstyle{bold orange}=[-, draw={rgb,255: red,255; green,128; blue,0}, fill=none, line width=1pt]
\tikzstyle{bold black}=[-, line width=1pt, draw=black, fill=none, tikzit draw=black]
\tikzstyle{bold red}=[-, draw={rgb,255: red,191; green,0; blue,64}, fill=none, line width=1pt]
\tikzstyle{fill light green}=[-, fill={rgb,255: red,166; green,166; blue,112}, fill opacity=0.5, draw={rgb,255: red,121; green,121; blue,81}, line width=1pt]
\tikzstyle{new edge style 0}=[-, fill=yellow, fill opacity=0.5, draw={rgb,255: red,146; green,146; blue,0}, tikzit fill=yellow, tikzit draw={rgb,255: red,146; green,146; blue,0}]
\tikzstyle{bold dashed red}=[-, draw={rgb,255: red,191; green,0; blue,64}, fill=none, line width=1pt, dashed]
\tikzstyle{bold dashed orange}=[-, draw={rgb,255: red,255; green,128; blue,0}, dashed, line width=1pt]
\tikzstyle{bold blue}=[-, draw={rgb,255: red,0; green,0; blue,228}, line width=1pt]
\tikzstyle{arrow red}=[draw={rgb,255: red,191; green,0; blue,64}, ->, line width=1pt]
\tikzstyle{new edge style 2}=[-, draw={rgb,255: red,191; green,0; blue,64}, line width=1pt]
\tikzstyle{boldish}=[-, line width=0.6mm, fill=cyan]
\tikzstyle{white edge}=[-, draw=white]
\tikzstyle{purple edge}=[-, draw={rgb,255: red,128; green,0; blue,128}, line width=1pt]
\tikzstyle{light gray}=[-, fill={rgb,255: red,191; green,191; blue,191}, draw=white, tikzit fill={rgb,255: red,191; green,191; blue,191}, tikzit draw={rgb,255: red,191; green,191; blue,191}, fill opacity=0.3]
\tikzstyle{invisible edge}=[-, fill opacity=0, fill=none]
\tikzstyle{carta da zucchero thin}=[-, style=fill carta da zucchero, line width=0.1pt, fill={rgb,255: red,129; green,158; blue,219}, tikzit fill={rgb,255: red,129; green,158; blue,219}]
\tikzstyle{pink thin}=[-, style=fill pink, line width=0.1pt, fill={rgb,255: red,193; green,100; blue,94}]
\tikzstyle{fill green thin edge}=[-, style=fill green, tikzit fill={rgb,255: red,62; green,128; blue,120}, line width=0.1pt]

\usepackage{xifthen}

\newcommand{\opapp}[2]{\ensuremath{#1\left(#2\right)}} %

\newcommand{\reals}{\ensuremath{\mathbb{R}}}

\newcommand{\tsuchthat}[2]{\ensuremath{\left\{#1\middle|#2\right\}}} %
\newcommand{\suchthat}[2]{\tsuchthat{\,#1\,}{\,#2\,}} %
\newcommand{\bra}[1]{\ensuremath{\left\langle#1\right|}}
\newcommand{\ket}[1]{\ensuremath{\left|#1\right\rangle}}

\newcommand{\downset}[1]{\ensuremath{#1\!\downarrow}}

\newcommand{\cprob}[2]{\opapp{\mathbb{P}}{#1\middle|#2}}

\newcommand{\restrict}[2]{#1|_{#2}}
\newcommand{\Subsets}[1]{\opapp{\mathcal{P}\!}{#1}} %

\newcommand{\ev}[1]{\text{#1}} %

\newcommand{\LsetsSym}{\Lambda} %
\newcommand{\Lsets}[1]{\opapp{\LsetsSym}{#1}} %

\newcommand{\DistSym}{\mathcal{D}}
\newcommand{\Dist}[1]{\opapp{\DistSym}{#1}}

\newcommand{\hist}[1]{%
    \ensuremath{
        \left\{
            \foreach \i\j [count=\idx] in {#1}{%
                \ifnum\idx=1%
                    \ev{\i}\!:\!\j%
                \else%
                    ,\,\ev{\i}\!:\!\j%
                \fi%
            }
        \right\}
    }
}
\newcommand{\evset}[1]{%
    \ensuremath{
        \left\{
            \foreach \i [count=\idx] in {#1}{%
                \ifnum\idx=1%
                    \ev{\i}%
                \else%
                    ,\ev{\i}%
                \fi%
            }
        \right\}
    }
}

\newcommand{\eqdef}{\stackrel{def}{=}}

\newcommand{\defn}{\textbf}
\renewcommand{\vec}[1]{{\underline{#1}}}

\newcommand{\FunType}[2]{#1 \rightarrow #2}
\newcommand{\JFunType}[1]{\FunType{I_{#1}}{O_{#1}}}
\newcommand{\CFunType}[1]{I_{#1} \stackrel{#1}{\longrightarrow} O_{#1}}
\newcommand{\JDistType}[1]{\FunType{I_{#1}}{\Dist{O_{#1}}}}
\newcommand{\CDistType}[1]{I_{#1} \stackrel{#1}{\longrightarrow} \Dist{O_{#1}}}

\begin{document}

\title{Local fraction in Static Causal Orders}
\author{Stefano Gogioso}
\email{stefano.gogioso@cs.ox.ac.uk}
\affiliation{University of Oxford}
\author{Nicola Pinzani}
\email{nicola.pinzani@ulb.be}
\affiliation{Université Libre de Bruxelles}

\begin{abstract}
    In this Letter, we introduce a notion of local fraction for experiments taking place against arbitrary static causal backgrounds---greatly generalising previous results on no-signalling scenarios \cite{elitzur1992quantum,barrett2006maximally,aolita2012fully,abramsky2017contextual}---and we explicitly formulate a linear program to compute this quantity.
    We derive a free characterization of causal functions which allows us to efficiently construct the matrices required to perform concrete calculations.
    We demonstrate our techniques by analysing the local fraction of a novel example involving two Bell tests in interleaved causal order.
\end{abstract}
\maketitle

\section{Introduction}
\label{sec:intro}

It is a fundamental prediction of quantum theory---dating all the way back to Bell's seminal 1964 work \cite{bell1964on}---that certain experimental setups are incompatible with the existence of a classical causal mechanism explaining the measurement statistics.
More specifically, the issue arises from the desire to localize the classical explanatory variables for any correlations to the common past of the relevant agents' interventions, according to the dictates of relativistic causality.
There is a tension between the causal structure imposed on the experimental devices and that required by a local classical explanation of the observed probability distributions on measurement outcomes, conditional on experimental parameters which are local to the agents and presumed to be freely chosen.
This tension takes the concrete form of quantum non-locality, a fundamental obstacle to the recovery of any classical explanatory model bound by the same no-signalling constraints as the quantum mechanical experiment.
The statistical nature of quantum theory is hence revealed to be radically different from that presumed by classical causal modelling \cite{pearl2009causality}: for an observer bound by the laws of causality, quantum randomness is intrinsic, and it cannot be reduced to mere ignorance of some latent, or ``hidden'', classical variable.

Quantum non-locality---as well as the broader notion of quantum contextuality---has been extensively studied as a resource in a variety of computational tasks.
Examples include measurement-based quantum computing \cite{howard2014quantum,frembs2018contextuality}, the study of quantum cryptographic protocols \cite{barrett2013memory,hillery1999quantum,vazirani2014fully,vazirani2016erratum}, and the formulation of an unconditional separation between classical and quantum circuits with shallow depth and bounded connectivity\cite{quantum2018bravyi,bravyi2020quantum,Bravyi2019QuantumAW}.
The quantification of non-locality and contextuality is thus a task of significant interest, and it has received significant attention in recent decades \cite{bancal2009quantifying,de2014on,abramsky2017contextual,wolfe2020quantifying}.
It is somewhat surprising, however, that such quantification efforts have largely remained confined to ``discrete'' scenarios, involving either a single system (contextuality) or a collection of spacelike separated agents (non-locality).

In this Letter, we generalise the ``local fraction'' metric \cite{abramsky2017contextual} from the discrete case investigated thus far to arbitrary static causal orders.
The metric was originally introduced by Abramsky and collaborators as part of the sheaf-theoretic framework for non-locality and contextuality \cite{abramsky2011sheaf,abramsky2014operational,abramsky2015contextuality,barbosa2019continuous,mansfield2014extendability}.
In doing this, we provide concrete tools---both mathematical and computational---for the quantification of non-locality as a resource in diverse settings, such as the formulation of novel device-independent cryptographic protocols, the characterization of quantum computational advantage, or the certification of foundational experiments on quantum causal structure.

\section{Operational Assumptions}

In this work, we study the non-locality of experiments and protocols set against a static causal background.
Specifically, by a \defn{static causal order} $\Omega = (|\Omega|, \leq)$ we will mean a set of events $|\Omega|$ equipped with a partial order relation $\leq$, subject to the following interpretation for two distinct events $\omega, \xi \in \Omega$: $\omega$ \defn{causally precedes} $\xi$ if $\omega < \xi$, it \defn{causally succeeds} $\xi$ if $\xi < \omega$, and the two events are \defn{causally unrelated} if $\omega \not\leq \xi$ and $\xi \not \leq \omega$.

Operationally, we interpret an individual event $\omega$ as the local operation of a black-box device in the context of the experiment: an input to the device is freely chosen (from a finite input set $I_\omega$), in response to which the device produces an output (probabilistically sampled from a finite output set $O_\omega$).
The probability distribution $\cprob{\vec{o}}{\vec{i}}$ on joint outputs $\vec{o} \in O_\Omega$ for all devices, conditional to joint inputs $\vec{i} \in I_\Omega$ for all devices, is the object of our non-locality analysis.
\begin{equation}
    I_\Omega
    \eqdef \prod_{\omega \in \Omega}I_\omega
    \hspace{1.5cm}
    O_\Omega
    \eqdef \prod_{\omega \in \Omega}O_\omega
\end{equation}
When we say that the devices are operated locally at each event, we mean that no information about the other events is explicitly used in the operation: every dependence on the inputs and outputs at other events must be entirely mediated by the causal structure.

\section{Causal Functions}

Let $\Omega$ be a static causal order, and consider the set of functions $f$ which associate a joint output $f\left(\vec{i}\right) \in O_\Omega$ to each joint input $\vec{i} \in I_\Omega$:
\begin{equation}
    \JFunType{\Omega}
    \hspace{1mm}
    \eqdef \left\{f: I_\Omega \rightarrow O_\Omega\right\}
\end{equation}
We wish to say that one such function $f$ is ``causal'' for $\Omega$ if it respects the constraints of ``no-signalling from the future'', stating that the output at an event $\xi$ is independent of the input at all events $\omega$ such that $\omega \not\leq \xi$, i.e., all events which don't causally precede $\xi$.
We will adopt an equivalent---but more structurally insightful---formulation of the no-signalling constraints, based on the order structure of $\Omega$.

We start by defining a \defn{lowerset} for $\Omega$ to be a subset $U \subseteq \Omega$ which is causally closed in the past, i.e., one such that for each event $\omega \in U$ all events $\xi < \omega$ which causally precede $\omega$ are also in $U$.
We write $\Lsets{\Omega}$ for the set of lowersets of $\Omega$, and we note in passing that $\Lsets{\Omega}$ endows the set $|\Omega|$ with the structure of a topological space.
A special case of lowerset is given by the causal past of each individual event $\omega \in \Omega$: this is known as its \defn{downset} $\downset{\omega}$, and it consists of all events $\xi \in \Omega$ such that $\xi \leq \omega$.

Given one such lowerset $U \in \Lsets{\Omega}$, we now define the \defn{restriction} $\restrict{f}{U}$ of a function $f: \JFunType{\Omega}$ to $U$ by restriction of its join input/output pairs:
\begin{equation}
    \restrict{f}{U} \eqdef \restrict{\vec{i}}{U} \mapsto \restrict{f(\vec{i})}{U}
\end{equation}
In general, the resulting restriction $\restrict{f}{U}$ is a one-to-many relation, rather than a well-defined function: for two distinct $\vec{i}, \vec{i'} \in I_\Omega$, it is possible to have $\restrict{\vec{i}}{U} = \restrict{\vec{i'}}{U}$ but $\restrict{f(\vec{i})}{U} \neq \restrict{f(\vec{i'})}{U}$.

An equivalent way to define causality is then to say that $f: \JFunType{\Omega}$ is \defn{causal} for $\Omega$ if its restrictions $\restrict{f}{U}$ to all lowersets of $\Omega$ are well-defined functions.
More explicitly, the \defn{no-signalling constraints} can be formulated in terms restrictions of input/output pairs:
\begin{equation}
    \restrict{\vec{i}}{U} = \restrict{\vec{i'}}{U}
    \;\Rightarrow\;
    \restrict{f(\vec{i})}{U} = \restrict{f(\vec{i'})}{U}
\end{equation}
We adopt the following notation for the set of causal functions on $\Omega$:
\begin{equation}
    \CFunType{\Omega}
    \hspace{2mm}
    \eqdef \suchthat{f: \JFunType{\Omega}}{f \text{ causal for }\Omega}
\end{equation}

Previous literature about local fraction in no-signalling scenarios deals with the special case where $\Omega$ is the discrete causal order, in which distinct events are causally unrelated.
In the discrete case, the lowersets are all subsets $\Lsets{\Omega} = \Subsets{\Omega}$, and causal functions are those whose restrictions $\restrict{f}{\{\omega\}}$ are well-defined all the way down to the individual events $\omega \in \Omega$.

\section{Free Characterization of Causal Functions}
\label{subsec:causal-funs-free}

It is known that causal functions for the discrete causal order admit an alternative ``free'' characterization, as the product of an arbitrary family of local functions $f^{(\omega)}: I_\omega \rightarrow O_\omega$, describing the behavior of devices at the individual events $\omega \in \Omega$:
\begin{equation}
    f(\vec{i}) := \left(f^{(\omega)}(i_\omega)\right)_{\omega \in \Omega}
\end{equation}
Here, ``free'' is used in opposition to the previous characterization in terms of no-signalling constraints: in the ``constrained'' characterization, causal functions are implicitly selected out of a much larger set of functions; in the ``free'' characterization, causal functions are constructed in terms of an explicit parametrization.

A natural question then arises: can we provide a similarly free characterizations of the causal functions on an arbitrary static causal order $\Omega$?
We proceed to answer this in the positive: causal functions are freely characterized by the outcome they produce at each event, conditional to the inputs in the past of that event.

Because each downset $\downset{\omega}$ is in particular a lowerset, the restriction $\restrict{f}{\downset{\omega}}: \JFunType{\downset{\omega}}$ of a causal function $f$ to $\downset{\omega}$ is well-defined: this is the function which maps joint inputs for all events in the past of $\omega$ to joint outputs for those events.
If we further restrict the output of $\restrict{f}{\downset{\omega}}$ to $\omega$ alone, we obtain the following function:
\begin{equation}
    f^{(\omega)} \eqdef \left(\restrict{f}{\downset{\omega}}\right)_{\omega}: \FunType{I_{\downset{\omega}}}{O_\omega}.
\end{equation}
The key observation is now that the original function $f$ can be uniquely reconstructed from the family of restrictions $f^{(\omega)}$ on all events $\omega \in \Omega$, as follows:
\begin{equation}
    \label{eq:caus-fun-free-construction}
    f(\vec{i}) = \left(
        f^{(\omega)}(\restrict{\vec{i}}{\downset{\omega}})
    \right)_{\omega \in \Omega}
\end{equation}
Our free characterization of causal functions $f$ on $\Omega$ then takes the form of a bijection with all families $\left(f^{(\omega)}\right)_{\omega \in \Omega}$ of functions $f^{(\omega)}$ specifying outputs for each event $\omega \in \Omega$:
\begin{equation}
    \CFunType{\Omega}
    \hspace{1mm}\cong
    \prod\limits_{\omega \in \Omega} \left(\FunType{I_{\downset{\omega}}}{O_\omega}\right)
\end{equation}

If the output sets are the same $O_\omega = \tilde{O}$ for all events, the free characterization above can be re-arranged into a simpler form
\footnote{
    This is because of the ``universal property'' of the coproduct $\coprod$, which is also known as the disjoint union: functions $F: \left(\coprod_{x \in X} A_x\right) \rightarrow B$ correspond exactly to families of functions $\vec{G} \in \prod_{x \in X} \left(A_x \rightarrow B\right)$.
    The correspondence sends $F \mapsto \left(G_x \eqdef \restrict{F}{A_x}\right)_{x \in X}$ in one direction, and $F$ can be uniquely reconstructed from the resulting family $\vec{G}$ as $F(a) = G_{x_a}(a)$, where $x_a$ is the unique index such that $a \in A_{x_a}$ (unique by disjointness of the union).
},
showing that causal functions on $\Omega$ are in bijection with all possible functions $F$ which associate an output value $F\left(\restrict{\vec{i}}{\downset{\omega}}\right) \in \tilde{O}$ to each possible \defn{input history} $\restrict{\vec{i}}{\downset{\omega}} \in I_{\downset{\omega}}$ for each event $\omega \in \Omega$:
\begin{equation}
    \CFunType{\Omega}
    \hspace{1mm}\cong
    \left(\coprod_{\omega \in \Omega} I_{\downset{\omega}}\right) \rightarrow \tilde{O}
\end{equation}
Input histories form the basis for a further generalization of non-locality and contextuality from static to dynamic and indefinite causal orders, as detailed in \cite{gogioso2022combinatorics,gogioso2022topology,gogioso2022geometry}.

\section{Causal Conditional Distributions}

We write $\Dist{X}$ for the space of probability distributions on a finite set $X$:
\begin{equation}
    \Dist{X} \eqdef
    \suchthat{
        d: X \rightarrow \reals^+
    }{
        \sum_{d \in X} d_x = 1
    }
\end{equation}
For a static causal order $\Omega$, we consider the set of probability distributions on joint outputs for all events, conditional to joint inputs:
\begin{equation}
    \JDistType{\Omega}
\end{equation}
The traditional notation for conditional probability can be defined in terms of conditional distributions, as follows:
\begin{equation}
    \mathbb{P}(\vec{o}|\vec{i}) \equiv d(\vec{i})_{\vec{o}}
\end{equation}

In analogy to causal functions, we define the \defn{restriction} of a conditional distribution $d: \JDistType{\Omega}$ to a lowerset $U \in \Lsets{\Omega}$ by restricting pairs of joint inputs and the corresponding distributions on joint outputs:
\begin{equation}
    \restrict{d}{U}
    \eqdef
    \restrict{\vec{i}}{U}
    \mapsto \restrict{d(\vec{i})}{U}
\end{equation}
The restricted probability distribution $\restrict{d(\vec{i})}{U} \in \Dist{O_U}$ is obtained by marginalisation:
\begin{equation}
    \restrict{d(\vec{i})}{U}
    \eqdef\left(
        \sum_{
            \substack{
                \vec{o} \in O_\Omega \\
                \restrict{\vec{o}}{U} = \vec{o'}
            }
        }
        d(\vec{i})_{\vec{o}}
    \right)_{\vec{o'} \in O_U}
\end{equation}
We then say that a conditional probability distribution $d: \JDistType{\Omega}$ is \defn{causal} for $\Omega$ if its restrictions $\restrict{d}{U}$ to all lowersets of $\Omega$ are well-defined conditional probability distributions.

More explicitly, the \defn{no-signalling constraints} can be formulated in terms of marginalisation:
\begin{equation}
    \restrict{\vec{i}}{U} = \restrict{\vec{i'}}{U}
    \;\Rightarrow\;
    \restrict{d(\vec{i})}{U} = \restrict{d(\vec{i'})}{U}
\end{equation}
Finally, we adopt the following notation for the set of causal distributions on $\Omega$:
\begin{equation}
    \CDistType{\Omega}
    \eqdef \suchthat{d: \JDistType{\Omega}}{d \text{ causal for }\Omega}
\end{equation}

Functions $f: \JFunType{\Omega}$ can be embedded into conditional distributions $\delta^f: \JDistType{\Omega}$:
\begin{equation}
    \delta^f \eqdef \vec{i} \mapsto \delta^{f(\vec{i})}
    \hspace{1cm}
    \delta^{f(\vec{i})}_{\vec{o}} \eqdef \begin{cases}
        1 &\text{ if } \vec{o} = f(\vec{i}) \\
        0 &\text{ otherwise}
    \end{cases}
\end{equation}
Causality of functions then arises as a special case of causality of distributions.

\section{Local Fraction}

Given a causal conditional distribution $d: \CDistType{\Omega}$, it is interesting to ask how much of the observed randomness can be explained ``locally and classically'', i.e., by a local deterministic hidden variable model.

For a causal conditional distribution $d: \CDistType{\Omega}$, a \defn{deterministic hidden variable model (DHVM)} is a decomposition of $d$ into a convex combination of functions $f: \JFunType{\Omega}$, where $\lambda \in \Dist{\JFunType{\Omega}}$ is some probability distribution over the set of functions:
\begin{equation}
    d =
    \hspace{-3mm} \sum_{f: \JFunType{\Omega}} \hspace{-4mm}
    \lambda_f \delta^f
\end{equation}
A \defn{local} DHVM is one where the functions appearing in the decomposition are constrained to be causal for the static causal order $\Omega$, i.e., one where $\lambda \in \Dist{\CFunType{\Omega}}$:
\begin{equation}
    d =
    \hspace{-3mm} \sum_{f: \CFunType{\Omega}} \hspace{-4mm}
    \lambda_f \delta^f
\end{equation}

Whether a causal conditional distribution $d: \CDistType{\Omega}$ admits a local DHVM is a yes/no question of great foundational interest, but with limited practical applicability.
Instead, we consider the \defn{local fraction} of $d$, a continuous metric defined as the largest mass $\mu \in [0,1]$ of $d$ which is explained by some local DHVM:
\begin{equation}
    \label{eq:local-fraction}
    \max \suchthat{
        \mu \in [0,1]
    }{
        \exists
        \lambda
        \text{ s.t. }
            \mu \hspace{-3mm} \sum_{f: \CFunType{\Omega}} \hspace{-4mm}
            \lambda_f \delta^f
        \leq d
    }
\end{equation}

We can also consider local DHVMs where the causal functions are restricted to certain sub-classes of interest.
For example, we define the \defn{no-signalling local fraction} of $d$ to be the largest mass which is explained by a local DHVM where $\lambda \in \Dist{\CFunType{\Omega_{disc}}}$ is restricted to the discrete causal order $\Omega_{ns} = (|\Omega|, \text{id}_\Omega)$ on the same set of events as $\Omega$.

\subsection*{Linear Program}

The following linear program, with variables $x_f \in \reals$ indexed by causal functions $f \in \CFunType{\Omega}$, can be used to compute the local fraction for a given causal conditional distribution $d: \CDistType{\Omega}$:
\begin{align*}
    \text{maximise } \sum_f x_f & \text{ subject to:}\\
    \forall f.\; x_f &\geq 0 \\
    \sum_f x_f \delta^f &\leq d
\end{align*}
With reference to our definition of local fraction, we have $\mu = \sum_f x_f$ and $\lambda_f = \sfrac{x_f}{\mu}$.

\section{Temporally Interleaved Bell Tests}

\begin{figure}[h!]
    \begin{center}
        \scalebox{1.2}{\begin{tikzpicture}[scale=0.8]
	\begin{pgfonlayer}{nodelayer}
		\node [style=none] (140) at (-2.75, -1.25) {};
		\node [style=none] (141) at (-3.5, -1.25) {};
		\node [style=none] (142) at (-3.5, -2) {};
		\node [style=none] (143) at (-2.75, -2) {};
		\node [style=none] (130) at (2.75, -1.25) {};
		\node [style=none] (131) at (3.5, -1.25) {};
		\node [style=none] (132) at (3.5, -2) {};
		\node [style=none] (133) at (2.75, -2) {};
		\node [style=none] (116) at (2.75, 1.25) {};
		\node [style=none] (117) at (3.5, 1.25) {};
		\node [style=none] (118) at (3.5, 0.5) {};
		\node [style=none] (119) at (2.75, 0.5) {};
		\node [style=eslabel] (0) at (-2.5, 1.5) {$C$};
		\node [style=none] (1) at (2.5, 1.5) {};
		\node [style=eslabel] (2) at (-2.5, -2.25) {$A$};
		\node [style=eslabel] (3) at (2.5, -2.25) {$D$};
		\node [style=wide point, line join = round] (4) at (-2, -4.25) {$\Phi$};
		\node [style=wide point, line join = round] (5) at (2, -4.25) {$\Phi$};
		\node [style=none] (8) at (1.25, -4) {};
		\node [style=none] (9) at (2.5, -4) {};
		\node [style=none] (10) at (-2.5, -4) {};
		\node [style=none] (11) at (-1.25, -4) {};
		\node [style=none] (40) at (-2.75, 1.25) {};
		\node [style=none] (41) at (-3.5, 1.25) {};
		\node [style=eslabel] (51) at (2.5, 1.5) {$B$};
		\node [style=none] (67) at (-3.5, 2) {};
		\node [style=none] (68) at (-4.75, 2) {};
		\node [style=none] (69) at (-4.75, 0.875) {};
		\node [style=none] (70) at (-3.5, 0.875) {};
		\node [style=none] (71) at (-2.75, 0.875) {};
		\node [style=none] (72) at (3.5, 2) {};
		\node [style=none] (73) at (4.75, 2) {};
		\node [style=none] (74) at (4.75, 0.875) {};
		\node [style=none] (75) at (3.5, 0.875) {};
		\node [style=none] (76) at (2.75, 0.875) {};
		\node [style=none] (77) at (-4.75, -1.625) {};
		\node [style=none] (79) at (-2.75, -1.625) {};
		\node [style=none] (80) at (4.75, -1.625) {};
		\node [style=none] (82) at (2.75, -1.625) {};
		\node [style=none] (83) at (-4.75, -2.75) {};
		\node [style=none] (84) at (-3.5, -2.75) {};
		\node [style=none] (85) at (4.75, -2.75) {};
		\node [style=none] (86) at (3.5, -2.75) {};
		\node [style=none] (87) at (-5.5, -1.625) {$o_A$};
		\node [style=none] (88) at (-5.5, -2.75) {$i_A$};
		\node [style=none] (89) at (-5.5, 0.875) {$i_C$};
		\node [style=none] (90) at (-5.5, 2) {$o_C$};
		\node [style=none] (91) at (5.5, 0.875) {$i_B$};
		\node [style=none] (92) at (5.5, 2) {$o_B$};
		\node [style=none] (93) at (5.5, -2.75) {$i_D$};
		\node [style=none] (94) at (5.5, -1.625) {$o_D$};
		\node [style=none] (95) at (-3.15, 0.875) {$\oplus$};
		\node [style=black dot] (96) at (-3.15, -1.625) {};
		\node [style=none] (97) at (3.1, 0.875) {$\oplus$};
		\node [style=black dot] (98) at (3.1, -1.625) {};
		\node [style=none] (99) at (-3.15, -0.25) {};
		\node [style=none] (101) at (3.1, -0.25) {};
		\node [style=none] (102) at (-3.5, 2.5) {};
		\node [style=none] (103) at (-1.5, 0.5) {};
		\node [style=none] (104) at (-1.5, 2.5) {};
		\node [style=none] (105) at (-2.75, 1.25) {};
		\node [style=none] (106) at (-3.5, 1.25) {};
		\node [style=none] (107) at (-2.75, 0.5) {};
		\node [style=none] (108) at (-3.5, 0.5) {};
		\node [style=none] (109) at (-2.75, 0.5) {};
		\node [style=none] (110) at (3.5, 2.5) {};
		\node [style=none] (111) at (1.5, 0.5) {};
		\node [style=none] (112) at (1.5, 2.5) {};
		\node [style=none] (113) at (2.75, 1.25) {};
		\node [style=none] (114) at (3.5, 1.25) {};
		\node [style=none] (115) at (2.75, 0.5) {};
		\node [style=none] (120) at (2.75, -2) {};
		\node [style=none] (121) at (3.5, -2) {};
		\node [style=none] (123) at (2.75, -1.25) {};
		\node [style=none] (127) at (2.75, -2) {};
		\node [style=none] (128) at (3.5, -2) {};
		\node [style=none] (134) at (-3.5, -3.25) {};
		\node [style=none] (135) at (-1.5, -1.25) {};
		\node [style=none] (136) at (-1.5, -3.25) {};
		\node [style=none] (137) at (-2.75, -2) {};
		\node [style=none] (138) at (-3.5, -2) {};
		\node [style=none] (139) at (-2.75, -1.25) {};
		\node [style=none] (144) at (3.5, -3.25) {};
		\node [style=none] (145) at (1.5, -1.25) {};
		\node [style=none] (146) at (1.5, -3.25) {};
		\node [style=none] (147) at (2.75, -2) {};
		\node [style=none] (148) at (3.5, -2) {};
		\node [style=none] (149) at (2.75, -1.25) {};
	\end{pgfonlayer}
	\begin{pgfonlayer}{edgelayer}
		\draw [style=light gray] (142.center)
			 to (143.center)
			 to (140.center)
			 to (141.center)
			 to cycle;
		\draw [style=light gray] (132.center)
			 to (133.center)
			 to (130.center)
			 to (131.center)
			 to cycle;
		\draw [style=light gray] (118.center)
			 to (119.center)
			 to (116.center)
			 to (117.center)
			 to cycle;
		\draw [style=bold black] (10.center) to (2);
		\draw [style=bold black, in=-90, out=90] (11.center) to (1.center);
		\draw [style=bold black, in=-90, out=90] (8.center) to (0);
		\draw [style=bold black] (9.center) to (3);
		\draw [style=light gray] (108.center)
			 to (109.center)
			 to (40.center)
			 to (41.center)
			 to cycle;
		\draw [style=arrow edge black] (67.center) to (68.center);
		\draw [style=arrow edge black] (69.center) to (70.center);
		\draw (70.center) to (71.center);
		\draw [style=arrow edge black] (72.center) to (73.center);
		\draw [style=arrow edge black] (74.center) to (75.center);
		\draw [style=arrow edge black] (83.center) to (84.center);
		\draw [style=arrow edge black] (85.center) to (86.center);
		\draw (96) to (95.center);
		\draw (98) to (97.center);
		\draw [style=arrow edge black] (96) to (99.center);
		\draw [style=arrow edge black] (98) to (101.center);
		\draw [style=fill white, line join = round] (103.center)
			 to (107.center)
			 to (105.center)
			 to (106.center)
			 to (102.center)
			 to (104.center)
			 to cycle;
		\draw (106.center) to (105.center);
		\draw [style=light gray] (105.center) to (107.center);
		\draw (107.center) to (105.center);
		\draw [style=fill white, line join = round] (111.center)
			 to (115.center)
			 to (113.center)
			 to (114.center)
			 to (110.center)
			 to (112.center)
			 to cycle;
		\draw [style=light gray] (113.center) to (115.center);
		\draw (115.center) to (113.center);
		\draw (75.center) to (76.center);
		\draw [style=light gray] (123.center) to (120.center);
		\draw [style=light gray] (120.center) to (121.center);
		\draw [style=fill white, line join = round] (135.center)
			 to (139.center)
			 to (137.center)
			 to (138.center)
			 to (134.center)
			 to (136.center)
			 to cycle;
		\draw (138.center) to (137.center);
		\draw [style=light gray] (137.center) to (139.center);
		\draw (139.center) to (137.center);
		\draw [style=fill white, line join = round] (145.center)
			 to (149.center)
			 to (147.center)
			 to (148.center)
			 to (144.center)
			 to (146.center)
			 to cycle;
		\draw (148.center) to (147.center);
		\draw [style=light gray] (147.center) to (149.center);
		\draw (149.center) to (147.center);
		\draw [style=arrow edge black] (79.center) to (77.center);
		\draw [style=arrow edge black] (82.center) to (80.center);
	\end{pgfonlayer}
\end{tikzpicture}}
    \end{center}
    \caption{
        Two temporally interleaved Bell tests, where the measurement outcome for the first party in each test affects the measurement choice for the second party in the other test.
    }
    \label{fig:protocol}
\end{figure}
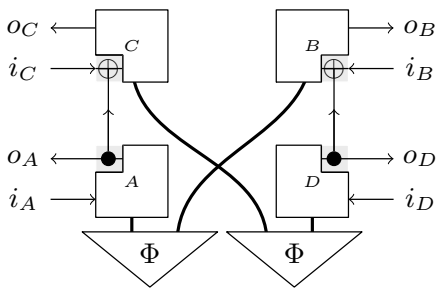

To demonstrate our techniques, we compute the local fraction for a novel example of temporally interleaved Bell tests, where two Bell states $\ket{\Phi^+}_{AB} \ket{\Phi^+}_{CD}$ are measured by four parties $\omega \in \{A, B, C, D\}$, with $A$ signalling to $C$ and $D$ signalling to $B$.
Each party measures their qubit in one of two bases in the ZY plane, at angles $\gamma_0, \gamma_1$ from the positive Z axis, and uses the measurement outcome as their output $o_\omega \in \{0,1\}$.
The measurement angles for $A$ and $D$ is determined directly by their inputs $i_A, i_D \in \{0,1 \}$, as $\gamma_{i_A}$ and $\gamma_{i_D}$ respectively.
The measurement angles for $C$ and $B$ are determined by the XOR of their respective inputs $i_C, i_B \in \{0,1 \}$ with the outputs $o_A, o_D \in \{0,1 \}$ of the party that preceded them, as $\gamma_{i_C \oplus o_A}$ and $\gamma_{i_B \oplus o_D}$ respectively.
Figure \ref{fig:protocol} displays the protocol in its entirety.

The probability distribution on joint outputs conditional to joint inputs factors as a product of two distributions, one for each pair of agents sharing a Bell state:
\begin{equation}
    \mathbb{P}(\vec{o}|\vec{i}) =
    \mathbb{P}(o_Ao_B|i_Ai_Bo_D)
    \mathbb{P}(o_Co_D|i_Ci_Do_A)
\end{equation}
The two distributions take the following form:
\begin{align}
    \mathbb{P}(o_Ao_B|i_Ai_Bo_D) &=
    \bra{\Phi^+}H_{i_A}^{o_A}\otimes H_{i_B \oplus o_D}^{o_B} \ket{\Phi^+}
    \\
    \mathbb{P}(o_Co_D|i_Ci_Do_A) &=
    \bra{\Phi^+}H_{i_C \oplus o_A}^{o_C}\otimes H_{i_D}^{o_D} \ket{\Phi^+}
\end{align}
where $H_{i}^{o}$ is the projector for outcome $o \in \{0, 1\}$ of a ZY plane measurement at angle $\gamma_i$:
\begin{equation}
    H_{i}^{o} \eqdef \frac{1}{2}\left(
        I + (-1)^o \left(\cos \gamma_{i} Z + \sin \gamma_{i} Y\right)
    \right)
\end{equation}
By construction, the conditional distribution is causal for the static causal order $\Omega$ defined below:
\begin{equation}
    \label{eq:protocol-caus-order}
    \Omega
    \hspace{3mm}\eqdef\hspace{2mm}
    \begin{array}{ccc}
        C && B \\
        \uparrow && \uparrow \\
        A && D
    \end{array}
\end{equation}
Figure \ref{fig:local-fraction} presents local fraction computations for this experimental setup, as a function of the experimental parameters $\gamma_0,\gamma_1 \in [0, \pi]$.
Figure \ref{fig:local-fraction-N} presents analogous local fraction computations for the case where we additionally allow signalling from $A$ to $B$.

\section{Discussion}
\label{sec:discussion}

\begin{figure*}
    \begin{center}
        \includegraphics[scale=0.45]{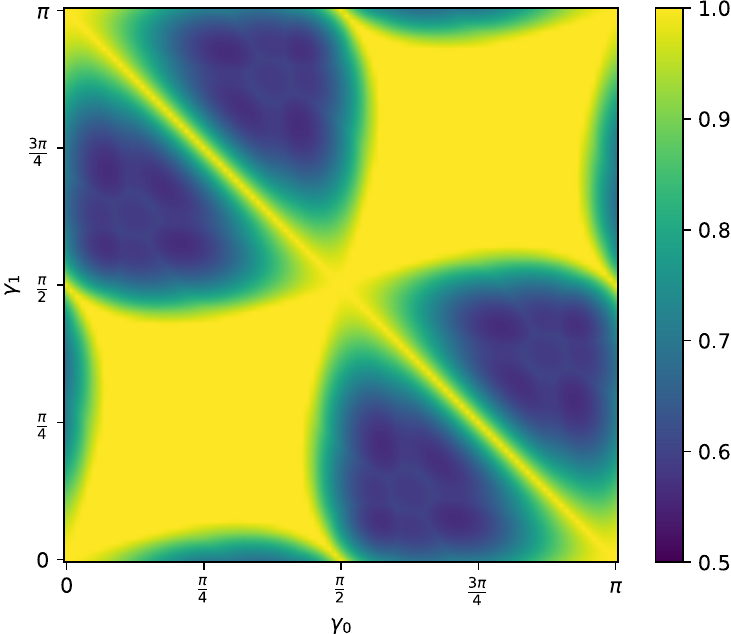}
        \includegraphics[scale=0.45]{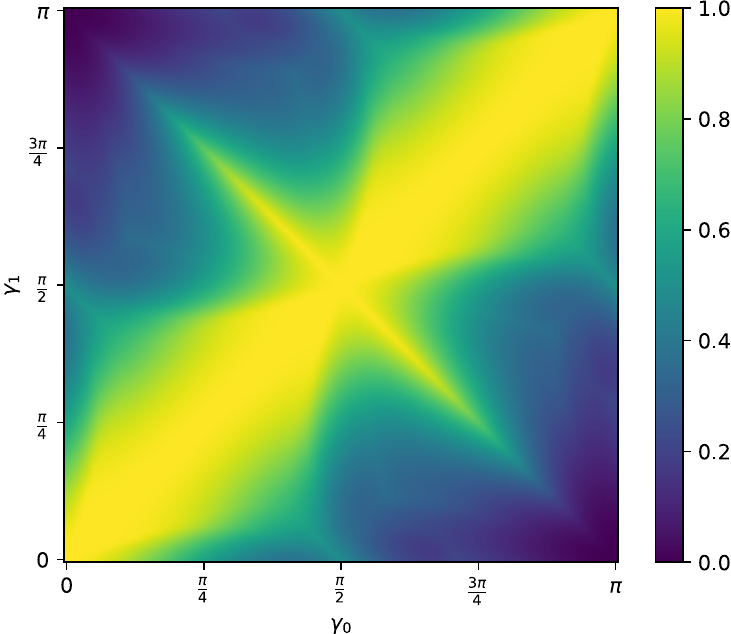}
        \includegraphics[scale=0.45]{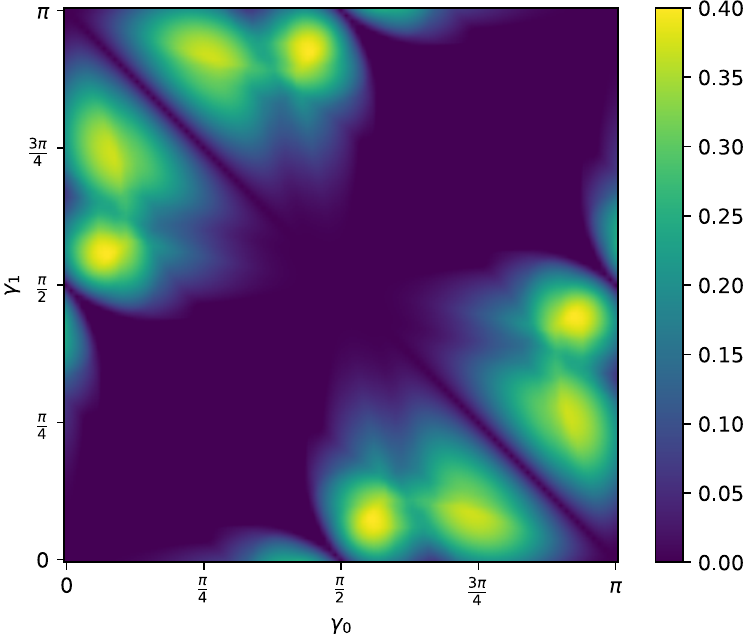}
    \end{center}
    \caption{
        Local fraction computations for the temporally interleaved Bell tests from Figure \ref{fig:protocol} as a function of the experimental parameters $\gamma_0,\gamma_1 \in [0, \pi]$, with respect to the causal order $\Omega$ from Equation \eqref{eq:protocol-caus-order}, where $A$ can only signal to $C$ and $D$ can only signal to $B$.
        Left: Local fraction.
        Middle: No-signalling local fraction.
        Right: Lower bound to the amount of signalling which cannot be explained classically.
    }
    \label{fig:local-fraction}
\end{figure*}

\begin{figure*}
    \begin{center}
        \includegraphics[scale=0.45]{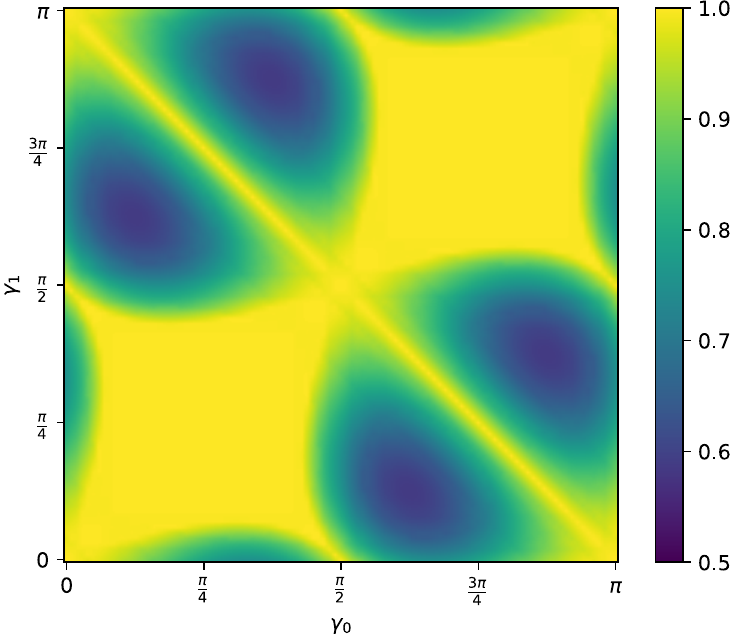}
        \includegraphics[scale=0.45]{non-signalling-explanations-interlaced-bell-res100.pdf}
        \includegraphics[scale=0.45]{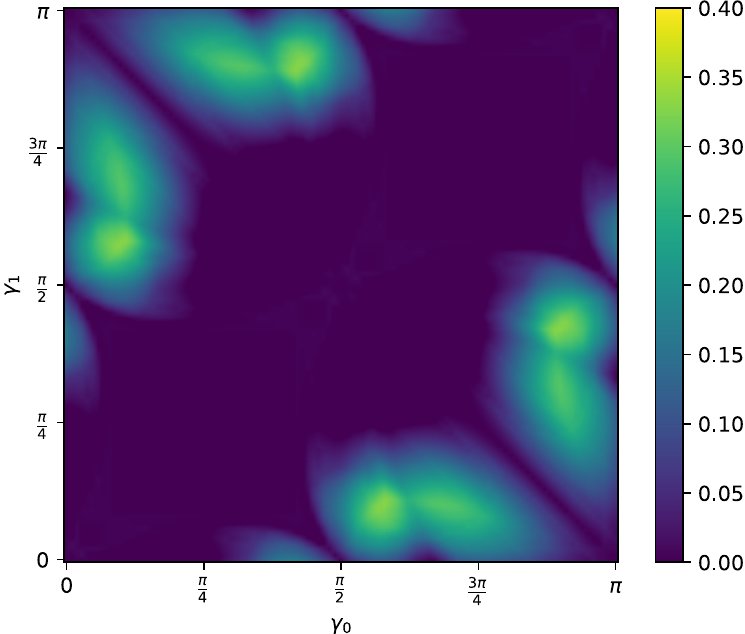}
    \end{center}
    \caption{
        Local fraction computations for the temporally interleaved Bell tests from Figure \ref{fig:protocol} as a function of the experimental parameters $\gamma_0,\gamma_1 \in [0, \pi]$, with respect to a causal order where $A$ can signal to both $B$ and $C$, while $D$ can only signal to $B$.
        Left: Local fraction.
        Middle: No-signalling local fraction.
        Right: Lower bound to the amount of signalling which cannot be explained classically.
    }
    \label{fig:local-fraction-N}
\end{figure*}

Until now, non-locality has been almost exclusively studied from the perspective of no-signalling scenarios: even within protocols and experiments with a non-trivial signalling structure, non-local correlations are often generated across causally unrelated events.
It may therefore be legitimate to ask whether the tools developed thus far, for the study of non-locality in no-signalling scenarios, could be sufficient to characterize non-locality in the presence of more general causal structures.
The first contribution of this work is to show that this is not the case.

We have presented a simple protocol where quantum correlations arise from two Bell scenarios, temporally interleaved by classical communication (cf. Figure \ref{fig:protocol}).
In Figure \ref{fig:local-fraction}, we show that the fraction of observed correlations which can be explained classically with respect to the underlying causal order $\Omega$ (left plot) is higher than the fraction which can be explained by techniques from previous literature, which are limited to the no-signalling causal order (middle plot).

As is the case in many other protocols and experiments, signalling in the protocol from Figure \ref{fig:protocol} is mediated by classical communication channels.
In such scenarios, it may be legitimate to ask whether non-classicality fully resides in the non-signalling fraction of the correlations, with all signalling effects classically explainable.
The second contribution of this work is to show that this is also not the case.

By combining local fraction, no-signalling local fraction and no-signalling fraction \cite{vallee2023corrected,lo2024developments}, Figure \ref{fig:local-fraction} demonstrates the existence of a non-trivial lower bound to the amount of signalling in our protocol which cannot be explained classically (right plot).
Specifically, we obtain the lower bound as follows: starting from 100\%, we subtract the local fraction and the no-signalling fraction, and then we add back the no-signalling local fraction (because the no-signalling local contribution was subtracted twice).
This is a lower bound, rather than an exact value, because the sub-distributions which maximise locality need not maximise no-signalling, and vice versa.
While this result might at first appear perplexing, it is explained by the interleaved structure of our protocol: non-classical information, in the form of two qubit channels, travels across the foliation separating events $A$ and $D$ in the past from events $C$ and $B$ in the future.

Finally, Figure \ref{fig:local-fraction-N} demonstrates that a non-trivial lower bound to the non-local signalling fraction (right plot) persists even in the case where one-directional signalling is allowed across the events of one of the two Bell tests.
In this new setting, the landscape of residual local fraction (left plot) coincides with that of a single Bell test, while the no-signalling local fraction (middle plot) is the same as the one for our original setting (the same no-signalling causal functions are common to both orders).

\section*{Acknowledgements}

This publication was made possible through the domain of the ID\# 61466 grant and ID\# 62312 grant from the John Templeton Foundation, as part of the project \href{https://www.templeton.org/grant/the-quantum-information-structure-of-spacetime-qiss-second-phase}{`The Quantum Information Structure of Spacetime' (QISS)}.
The opinions expressed in this publication are those of the authors and do not necessarily reflect the views of the John Templeton Foundation.
This publication was supported by the Program of Concerted Research Actions (ARC) of the Université Libre de Bruxelles.
Financial support by Hashberg Ltd for computational experiments is gratefully acknowledged.

\bibliographystyle{apsrev4-1}
\bibliography{biblio}

\end{document}